\documentclass[technote, letterletter, final, 9pt]{IEEEtran}

\usepackage[dvips]{graphics}
\usepackage[dvips]{changebar}
\usepackage{amsmath,bm}
\usepackage{rotating}
\usepackage[strings]{underscore}
\usepackage{setspace}
\usepackage{times}
\usepackage{color}
\usepackage{soul}
\usepackage{amsmath} 
\usepackage{amssymb}
\usepackage[strings]{underscore}
\usepackage{graphics,color}
\usepackage{epstopdf}
\usepackage{graphicx}
\usepackage{multirow}
\usepackage{ifpdf}

\usepackage{subfig}

\newcommand{\mb}[1]{{\mathbf #1}}

\begin{document}

\setcounter{page}{1}
\newcounter{mytempeqncnt}
\setlength{\textfloatsep}{-1pt}
\setlength{\intextsep}{1pt}
\title{Low-Complexity and Basis-Free Channel Estimation for Switch-Based mmWave MIMO Systems via Matrix Completion}
\author{ Rui Hu, Jun Tong, Jiangtao Xi, Qinghua Guo and Yanguang Yu 
\thanks{
The authors are with the School of Electrical, Computer and Telecommunications Engineering, University of Wollongong, Wollongong, NSW 2522, Australia. Email:  rh546@uowmail.edu.au, \{jtong, jiangtao, qguo, yanguang\}@uow.edu.au.} 
} 

	\maketitle

\begin{abstract}
Recently, a switch-based hybrid massive MIMO structure that aims to reduce the hardware complexity and 
improve the energy efficiency has been proposed as a potential candidate for millimeter wave (mmWave) communications. Exploiting the sparse nature of the mmWave channel, compressive sensing (CS)-based channel estimators have been proposed.  
When applied to real mmWave channels, the CS-based channel estimators may encounter heavy computational burden due to the high dimensionality of the basis. 
Meanwhile, knowledge about the response of the antenna array, which is needed for constructing the basis of the CS estimators, may not be perfect due to array uncertainties such as phase mismatch among array elements. This can result in the loss of sparse representation and hence the degraded performance of the CS estimator. In this paper, we propose a novel matrix completion (MC)-based low-complexity channel estimator. The proposed scheme is compatible with switch-based hybrid structures, does not need to specify a basis, and can avoid the basis mismatch issue. Compared with the existing CS-based estimator, the proposed basis-free scheme is immune to array response mismatch and exhibits a significantly lower complexity. 
Furthermore, we evaluate the impact of channel estimation scheme on the achievable spectral efficiency (SE) with antenna selection. 
The numerical results demonstrate that the MC estimator can achieve SE close to that with perfect channel state information.   
\end{abstract}


\vspace{-3ex}
\section{Introduction}

The enormous amount of spectrum at millimeter wave (mmWave)  frequencies (30-300 GHz) and the development in mmWave devices manufacturing technologies make the mmWave communication an attractive candidate for the 5G 
cellular network \cite{Key elements 5G}. {Large-scale multiple-input multiple-output (MIMO) transmission is suggested for mmWave systems to 
compensate for the significant signal attenuation in the mmWave band. However, a fully digital transceiver structure incurs significant {power consumption} by the large amount of radio frequency (RF) chains. Phase shifters- or switches-based hybrid structures that employ only a few RF chains have generated considerable interests recently \cite{channel estimation 1}, \cite{switches or phase shifters}. } 

{Employing large-scale MIMO leads to a large 
channel matrix which needs to be estimated for designing precoders and detectors. Using a conventional channel estimator such as the least squares (LS) estimator demands a large amount of training resources. Fortunately, the mmWave channel matrix tends to be low-rank due to the poor scattering nature at mmWave frequencies \cite{Key elements 5G}, \cite{Channel Measure}. This sparse nature can be exploited to reduce the training data requirement. Compressive sensing (CS)-estimators have recently been proposed for phase shifter- \cite {channel estimation 2} and switch-based \cite {switches or phase shifters} mmWave systems, which can reduce the required training time. The CS-based estimators generally need first define and quantize a searching basis, and have good performance with the assumption that the antenna array in the system is ideal, i.e., the predefined basis in the CS-based method is perfectly matched with the actual physical model of the channel. 
However, in practice, there often exist uncertainties regarding the array response, e.g., due to gain and phase mismatch, mutual coupling and position errors \cite{directional finding partly calibrated}.
With such array uncertainties, it is challenging to construct a proper basis upon which the channel is sparse. Since the performance of a CS-based estimator is sensitive to the choice of the basis \cite{CS with coherent dic},  \cite{basis mismatch}, a mismatched basis can result in significant performance degradation. Furthermore, the CS methods such as orthogonal matching pursuit (OMP) \cite{switches or phase shifters} may suffer from heavy computational load when fine grids are applied to achieve good performance.} 

In this paper, we study the channel estimation problem for single user switch-based mmWave systems \cite{switches or phase shifters} and show the sensitivity of the existing OMP estimator \cite{switches or phase shifters} to the phase mismatch of the array. We also propose a basis-free matrix completion (MC)-based channel estimation scheme and show that the mmWave channel satisfies the incoherence properties that enable accurate recovery of the full channel matrix from only a subset of its entries that are sampled uniformly randomly \cite{the power of convex relaxation}. We then discuss 
a training scheme that involves only properly controlling the switches at the transmitter and the receiver, which is compatible with the targeted hybrid structure. This scheme guarantees a high probability that at least one sample from each column and each row of the channel matrix is obtained. The singular-value projection (SVP) algorithm \cite{SVP} is applied to implementing the MC-based estimator and its complexity and parameter choice are analyzed. The simulation results show that the MC scheme, which does not need to specify a basis, has lower complexity than the existing CS-based scheme \cite{switches or phase shifters} and is immune to the phase mismatch of the array.

The paper is organized as follows. We introduce the switch-based hybrid mmWave system and review a CS-based channel estimator in Section II. In Section III, we present the proposed MC-based channel estimator. We show the simulation results in Section IV and conclude the paper in Section V. 
\vspace{-4.5ex}
\section{System Model}
\vspace{-1.5ex}
We consider a single user downlink 
mmWave MIMO system {which is the same as in \cite{switches or phase shifters}}. The system employs the array-subarray hybrid structure (A6) of \cite{switches or phase shifters} at the mobile station (MS): 
 At the MS, each of the $N_{\rm MS}$ transmit antennas is equipped with a switch, and every ${N_{\rm MS}}/{N_{\rm RF_{\rm MS}}}$ neighbouring switches are grouped together and connected to one of the {${N_{\rm RF_{\rm MS}}}$} RF chains. 
The BS has the same structure, with {$N_{\rm BS}$ antennas and $N_{\rm RF_{\rm BS}}$ RF chains.} 
The diagram of this structure 
is shown in Fig. 1. 
\begin{figure}
\label{A6}
\includegraphics[width=\columnwidth]{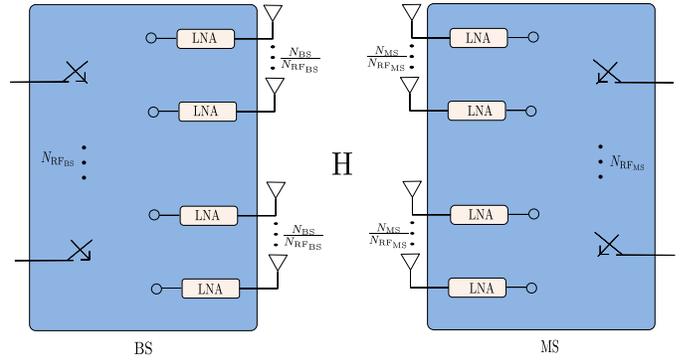}
\caption{Switch-based transmitter and receiver structure following the A6 structure of \cite{switches or phase shifters}, where LNA denotes low noise amplifier.}
\end{figure}
Following \cite{channel estimation 2}, the $N_{\rm MS}\times N_{\rm BS}$ downlink mmWave channel matrix is given by   
                             \begin{equation}
		               \label{mmWave}
		               		 \mb H = \mb A_{\rm MS}  \text{diag} (\bm{\alpha})  \mb A_{\rm BS}^H,
		          \end{equation}
where
 $\mb A_{\rm BS}=[\mb a_{\rm BS}(\phi_1), \mb a_{\rm BS} (\phi_2), \ldots, \mb a_{\rm BS}(\phi_L)]$, $(\cdot)^H$ denotes conjugate transpose, $\mb a_{\rm BS}(\phi_l)$ is a steering vector of the angle of departure (AoD) $\phi_l$  of the $l$-{th} path, and $L$ is the number of paths.  Similarly, we can define $\mb A_{\rm MS}=[\mb a_{\rm MS}(\theta_1), \mb a_{\rm MS}(\theta_2), \ldots, \mb a_{\rm BS}(\theta_L)]$, where $\mb a_{\rm MS}(\theta_l)$ is the steering vector of the angle of arrival (AoA) $\theta_l$.   
Assuming ideal uniform linear arrays (ULA) with distance $d$ between adjacent antennas and there are no amplitude, phase or antenna positioning errors, the steering vector is given by   
 \begin{equation} 
 \label{aBS} \mb a_{\rm BS}(\phi_l)=\frac{1}{\sqrt{N_{\rm BS}}}[1, \mathrm{e}^{j \frac{2\pi}{\lambda}d\sin(\phi_l)},
\cdots,\mathrm{e}^{j(N_{\rm BS}-1)\frac{2\pi}{\lambda}d\sin(\phi_l)} ]^T, 
\end{equation} 
where $\lambda$ is the wavelength. The steering vector $\mb a_{\rm MS}(\theta_l)$ is defined similarly.  
 The path gains are modeled as 
 \[  \bm{\alpha}=\sqrt{\frac{N_{\rm BS}N_{\rm MS}}{L}}[\alpha_1, \alpha_2, \ldots, \alpha_L]^T,\] 
where $\alpha_l$ is the complex gain of the $l$-th path, which is assumed to be i.i.d. $\mathcal{CN}(0,\sigma_\alpha^2)$ distributed.

The OMP can be applied to estimating the above $\mb H$ \cite{channel estimation 1}-\cite{channel estimation 2}, especially for channels with a small number $L$ of paths.
Ignoring the quantization error, using the virtual channel representation $\mb H$  may be modeled as \cite{switches or phase shifters}, \cite{compressed channel sensing}, \cite{virtual model},
 \begin{equation}
\label{CS model}
\mb H =\mb A_{\rm MSD}\mb H_v\mb A_{\rm BSD}^H, 
\end{equation}
where $\mb A_{\rm MSD}\in \mathbb{C}^{N_{\rm MS}\times{G_r}}$ and $\mb A_{\rm BSD}\in \mathbb{C}^{N_{\rm BS}\times{G_t}}$ are two dictionary matrices, and $\mb H_v\in\mathbb{C}^{G_r\times G_t}$ is a sparse matrix that contains the path gains of the quantized directions.  The two dictionary matrices  $\mb A_{\rm MSD}$ and $\mb A_{\rm BSD}$  are commonly constructed using steering vectors \cite{switches or phase shifters}, \cite{channel estimation 2}. Vectorizing (\ref{CS model}) leads to 
\begin{equation} 
\label{psi}
{\rm vec}(\mb H)=\bm \Psi \mb x, \quad \text{with}\hspace{0.2cm}\bm \Psi =  \mb A_{\rm BSD}^\ast \otimes \mb A_{\rm MSD} 
\end{equation}  
where $\mb \Psi$ is the basis matrix, $(\cdot)^\ast$ denotes conjugate, $\otimes$ represents Kronecker product, and  $\mb x \triangleq \mathrm{vec}(\mb H_v)$ is an $L$-sparse vector. Noisy observations of linear combinations of the entries of ${\rm vec}(\mb H)$  may be obtained by training, yielding 
\begin{equation}
\label{yvector} 
  {\mb y = \bm \Phi {\rm vec}(\mb H) + \mb z =  \bm \Phi \bm \Psi \mb x + \mb z,}
\end{equation}  
where $\bm \Phi$ is the sensing matrix specified by the training scheme and $\mb z$ is the noise. 
The OMP method finds the non-zero entries of $\mb x$ from $\mb y$, which corresponds to finding $L$ out of $G_r  G_t$ candidate direction pairs.
In order to obtain the row orthogonality of the two dictionaries, the physical angles of the steering vector should be generated according to the following equation \cite{channel estimation via OMP}
\begin{equation}
\label{dic_angle}
\frac{2{\pi}d}{\lambda}\sin (\theta_g)=\frac{2\pi}{G}(g-1)-\pi, g=1,2,\ldots, G,
\end{equation}
where $G$ is the number of grid points, $d$ is the distance between two neighbouring elements, and $\lambda$ is the wavelength. 
If $d=\frac{\lambda}{2}$, (\ref{dic_angle}) simplifies to 
\vspace{-1.5 ex}
\begin{equation}
\label{simplified}
\sin (\theta_g)=\frac{2}{G}(g-1)-1. 
\end{equation}
Under this condition, when the numbers of gird points $G_t=N_{\rm{BS}}, G_r=N_{\rm{MS}}$, the two dictionary matrices $\mb A_{\rm MSD}$ and $\mb A_{\rm BSD}$ are unitary. 
When  $G_t> N_{\rm{BS}}$ and $G_r> N_{\rm{MS}}$, the two dictionary matrices are redundant. The computational complexity of the OMP method is about $8MG_tG_r$ flops per iteration, where $M$ is the number of sampled entries. In general, the larger the numbers of grid points the better the performance, yet the heavier the computational burden. 

The above analysis of CS-based estimator is under the assumption that the channel has a sparse representation under the ideal steering vector basis. When uncertainties about the array response are presented as mentioned in Section I, the actual channel may not be sparse on the basis defined in (\ref{psi}).
Denote the unknown phase error at antenna element $i$ as $\gamma_i$. The real steering vector is
\begin{align}
\label{errorsteering}
\mb a_{\rm BS_{real}}(\phi_l)&=\frac{1}{\sqrt{N_{\rm BS}}}[\mathrm{e}^{j\gamma_1}, \mathrm{e}^{j (\frac{2\pi}{\lambda}d\sin(\phi_l)+\gamma_2)},\\ \nonumber
                                               &\cdots,\mathrm{e}^{j(\frac{2\pi}{\lambda}(N_{\rm BS}-1)d\sin(\phi_l)+\gamma_{N_{\rm{BS}}})}]^T, 
\end{align}
which depends not only on the AoA and AoD, but also on the phase error. In this paper, we will demonstrate the performance degradation of the OMP estimator due to phase mismatch through simulations in Section IV.
Even the unknown phase mismatch can cause the basis mismatch issue \cite{basis mismatch} on CS-based methods, the channel matrix $\mb H$ itself remains to be low-rank whenever the number of paths $L$ is small.
In Section III, we introduce a MC approach which is basis-free and is thus immune to the uncertainties of array response. 
\section{Matrix Completion for mmWave Channel Estimation}
The MC problem is to recover an unknown low-rank matrix $\mb M$ from a subset of entries sampled through the operator $P_{\Omega}(\cdot)$ defined by 
\vspace{-1.5ex}
\begin{equation}  
\label{operator}
[P_{\Omega}( \mb X)]_{i,j}=
      \begin{cases}
        [ \mb X]_{i,j},    & \quad (i,j) \in \Omega\\
          0,	   & \quad  \text{otherwise}
       \end{cases}, 
\end{equation} 
where $[\mb X]_{i,j}$ denotes the $(i, j)$-th entry of $\mb X$. The number of sampled entries of $\mb X$ in the operator $P_{\Omega}(\cdot)$ is $pN$, where $p$ is the sampling density and $N$ is the total number of entries in $\mb X$.
The recovery task is to solve
\vspace{-1ex}
\begin{equation}  
\min_{\mb X} {\rm rank}(\mb X), \quad \quad  \mathrm{s.t.} \quad  P_{\Omega}( \mb X)=P_{\Omega}( \mb M). 
\end{equation} 
This problem is NP-hard and usually solved approximately, e.g., as a nuclear norm minimization problem \cite{SVT} or an affine rank minimization problem (ARMP) \cite{SVP}. In this section, following the similar steps in \cite{array signal MC}, we first show the suitability of MC for mmWave channel estimation by examining the incoherence property of the mmWave MIMO channel. We then introduce a training scheme that is compatible with the switch-based structure and discuss the estimation algorithm and its complexity. 
\vspace{-1.5ex}


\subsection{Incoherence Property of mmWave Channel}
We assume large $N_{\rm MS}$ and $N_{\rm BS}$, which is of interest for mmWave applications. We first check the incoherence property of the mmWave channel with ideal antenna arrays.  
 Let the singular value decomposition (SVD) of the rank-$L$ matrix $\mb H$ be 
                      \begin{equation}
                       \label{matrix}
                      \mb H=\sum_{k=1}^L \sigma_k \mb u_k \mb v_k^H,   
                     \end{equation}
where $\sigma_k$ denotes the $k$-th singular value and $\mb u_k$ and $\mb v_k$ are the corresponding left and right singular vectors, respectively.
Define 
                          \begin{equation}
                                      \label{Uprojectopn}
                                       \mb P_U=\sum_{i=1}^L \mb u_i\mb u_i^H,\quad  \mb P_V=\sum_{i=1}^L \mb v_i\mb v_i^H, \quad  
                                       \mb E=\sum_{i=1}^L \mb u_i\mb v_i^H.
                                         \end{equation}
Let $\mb e_a$ denote the vector with the $a$-th entry equal to 1 and others equal to zero, $1_{a=a'}=1$ if $a=a'$ is true and $1_{a=a'}=0$ otherwise. If there exists $\mu$ such that  
\begin{itemize}

\item  for all pairs $(a,a')$ and $(b,b')$  
                   \begin{eqnarray}
                  \label{A11} 
                  \left| \langle \mb e_a, \mb P_U\mb e_{a'} \rangle -\frac{L}{N_{\rm MS}}1_{a=a'}\!\right|\!\!\!&\leq&\!\!\! \mu \frac{\sqrt{L}}{N_{\rm MS}} \\  \label{A12}
                  \left|\langle \mb e_b, \mb P_V\mb e_{b'} \rangle -\frac{L}{N_{\rm BS}}1_{b=b'}\!\right|\!\!\!&\leq&\!\!\! \mu \frac{\sqrt{L}}{N_{\rm BS}},  
                 \end{eqnarray}

\item and for all $(a,b)$,

                  \begin{equation}
                     \label{A2}
                     |\mb E_{ab}|\leq \mu \frac{\sqrt{L}}{\sqrt{N_{\rm MS} N_{\rm BS}}}, 
                     \end{equation}
\end{itemize}
then $\mb H$ obeys the strong incoherence property with parameter $\mu$ \cite{matrix completion with noise}. In this case,  $\mb H$ can be recovered without error with high probability if at least $C_1\mu^4n(\log n)^2$ uniformly sampled
entries are known \cite{matrix completion with noise}, where $C_1$ is a constant and $n=\max(N_{\rm MS}, N_{\rm BS})$.

We start examining the incoherence property of $\mb H$ from $L=1$. 
When $L=1$, comparing (\ref{mmWave}) with (\ref{matrix}), we can see that $\mb u_1=\mb a_{\rm MS}(\theta_1)$ and $\mb v_1=\mb a_{\rm BS}(\phi_1)$ are the singular vectors, all entries of $\mb P_U$ have the same module $ {1}/{N_{\rm MS}}$. 
When $a=a'$,   
\[ \langle \mb e_a, \mb P_U\mb e_{a'} \rangle =[\mb P_U]_{a, a} =\frac{1}{N_{\rm MS}},\] which yields $|\langle \mb e_a, \mb P_U\mb e_{a'} \rangle -\frac{1}{N_{\rm MS}}1_{a=a'}|=0$. 
When $a\neq a'$, 
\[ \left|  \langle \mb e_a, \mb P_U\mb e_{a'} \rangle \right|=|[\mb P_U]_{a, a'}| =\frac{1}{N_{\rm MS}}.\]  
We can now verify that (\ref{A11}) is satisfied with $\mu = 1$. Similarly, we can verify (\ref{A12}) and (\ref{A2}) with $\mu = 1$.   

For $L \geq 2$, we exploit the following asymptotic property of mmWave channel \cite{capacity steering}: As $N_{\rm MS}$ and $N_{\rm BS}$ become very large, the singular vectors of $\mb H$ converge to the steering vectors. Assume $\mb u_i = \mb a_{\rm MS}(\theta_i)$, $\mb v_i = \mb a_{\rm BS}(\phi_i)$, then all the entries of the left singular vectors have module $1/\sqrt{N_{\rm MS}}$ and those of the right singular vectors have module ${1}/\sqrt{N_{\rm BS}}$. Consequently, for $a=a'$, 
\begin{equation} 
\label{rge2eq1}  
\langle \mb e_a, \mb P_{U} \mb e_{a'} \rangle = [\mb P_U]_{a,a}=\frac{L}{N_{\rm MS}},
\end{equation}  
and for $a\ne a'$, 
\vspace{-2.5ex}
\begin{eqnarray}
\label{inequality1} \nonumber 
|\langle \mb e_a, \mb P_U\mb e_{a'} \rangle | \!& =  &\!  |[\mb P_{U}]_{a, a'}| = \left|\sum_{i=1}^{L} u_{i, a}u^\ast_{i,a'} \right | \\ &\le&   \sum_{i=1}^{L} |u_{i, a}||u^\ast_{i,a'}| = \frac{L}{N_{\rm MS}}
\end{eqnarray}
From (\ref{rge2eq1}) and (\ref{inequality1}) we can verify that (\ref{A11}) is satisfied with $\mu = \sqrt{L}$. 
Similarly we can verify (\ref{A12}) and (\ref{A2}). 

Based on the above analysis, the mmWave channel $\mb H$ without phase mismatches obeys the strong incoherence property with parameter $\mu\approx \sqrt{L}$ when $\mb H$ is large and thus can be recovered from a subset of its entries according to the MC theory.  

When phase mismatches are present,
it can be seen from (\ref{errorsteering}) that compared to the ideal steering vector, the amplitude of each element in the real steering vector $ \mb a_{\rm BS_{real}}(\phi_l)$ does not change. 
Hence the analysis for the channel under ideal antenna array assumption can still stand with the channel that has phase mismatch. 
The above analysis assumes noiseless samples of $\mb H$ and provides useful guidelines for high-SNR applications.  
\vspace{-2ex}
\subsection{Training Scheme}
The sampling pattern has a crucial influence on the performance of MC. 
From \cite{ matrix completion with noise}, at least one entry must be sampled from each row and each column to recover the original matrix. 
In this paper, we adapt the uniform spatial sampling (USS) scheme \cite{array signal MC}, which was proposed for array signal processing and seems to outperform alternative sampling schemes such as the Bernoulli scheme \cite[Section IV]{matrix completion with noise}. 

Suppose $M$ entries of the $N_{\rm MS}\times N_{\rm BS}$ matrix $\mb H$ need to be sampled.  The USS scheme suggests to take $ {M}/{N_{\rm BS}}$ distinct samples from the  $N_{\rm MS}$ entries of each column. In our switch-based array-subarray structure, there are $N_{\rm sub} \triangleq  {N_{\rm MS}}/{N_{\rm RF_{\rm_{MS}}}}$ antennas in each MS subarray which share the same RF chain. In order to make full use of the MS RF chains and keep the training time short, all the $N_{\rm RF_{\rm MS}}$ MS RF chains are activated during the whole training process. 
Each MS RF chain is switched randomly to a distinct antenna in the associated subarray and $N_{\rm RF_{\rm MS}}$ samples can be produced at each training stage. Thus, in total $ N_s = M/N_{\rm RF_{\rm MS}}$ training stages are used. 

We now describe the training process. 
We index the BS antennas from $1$ to $N_{\rm BS}$ and MS antennas from $1$ to $N_{\rm MS}$. 
Let $\mb Y \in \mathbb{C}^{N_{\rm MS} \times N_{\rm BS}}$ and initialize all its entries to zero. 
For each MS subarray $k$, $k=1,2,\cdots, N_{\rm RF_{\rm MS}}$, denote by $\mathcal{I}_k$ the set of the  antennas that have not been switched on so far. 
The disjoint sets $\{\mathcal{I}_k\}$ are initialized according to the array structure and the union of the initial $\mathcal{I}_k$ gives $\{1, 2, \cdots, N_{\rm MS}\}$. 
At the $t$-th training stage, 
\begin{itemize}
\item At the BS, only the transmit antenna indexed by $j_t \equiv {\rm mod} (t, N_s)$ is activated and a known symbol $s$ is sent. For each MS subarray $k$, randomly switch on (with equal probabilities) an antenna in $\mathcal{I}_k$ and 
denote by $i_k$ the index of the antenna switched on. 
The received symbol at the $i_k$-th MS antenna is written as   
 \vspace{-1ex}
\begin{equation}
  r_{i_k} =   [\mb H]_{i_k, j_t}  s +   n_{i_k},  \quad k=1,2, \cdots, N_{\rm RF_{\rm MS}}. 
\end{equation}
\item For $k=1,2,\cdots, N_{\rm RF_{\rm MS}}$,  

\[
	 [\mb Y]_{i_k, j_t} = \frac{  r_{i_k}}{s},    
\]

and remove $i_k$ from $\mathcal{I}_k$. 
\end{itemize}

The above simple training process yields noisy observations of $M$ distinct entries of $\mb H$, which are recorded in $\mb Y$.  
When $N_s \ge  N_{\rm BS}$, i.e., $M \ge N_{\rm BS} N_{\rm RF_{\rm MS}}$, it is guaranteed that at least one entry is observed (with noise) for each column of $\mb H$ as every BS antenna is switched on at least once. 
The sampling scheme also guarantees that for each MS subarray, $M/(N_{\rm BS} N_{\rm RF_{\rm MS}})$ out of the $N_{\rm sub}$ entries of each column of $\mb H$ have been sampled once. The event of missing an entire row of $\mb H$  corresponds to the case that for all the $N_{\rm BS}$ columns, the $M/(N_{\rm BS} N_{\rm RF_{\rm MS}})$ entries are taken from a common subset of the subarray with size $N_{\rm sub}-1$.  
The probability of such an event is  
\begin{equation}
\label{Puss}
P_{\rm miss} = \left(\frac{\left(\!
    \begin{array}{c}
      N_{\rm sub}-1 \\
      \frac{M}{N_{\rm BS}N_{\rm RF_{\rm MS}}}
    \end{array}
  \!\right)}{\left(\!
    \begin{array}{c}
      N_{\rm sub} \\
      \frac{M}{N_{\rm BS}N_{\rm RF_{\rm MS}}}
    \end{array}
  \!\right)}\right)^{N_{\rm BS}} \!\! \!\! \!\! \!\!\! =\left(\frac{N_{\rm MS}-\frac{M}{N_{\rm BS}}}{N_{\rm MS}}\right)^{N_{\rm BS}}, 
\end{equation}
which is negligible when $M$ and $N_{\rm BS}$ are large enough. 
For example, when $N_{\rm MS}=64$, $N_{\rm BS}=64$, and $M=0.5 \times N_{\rm MS}  N_{\rm BS}$, 
$P_{\rm miss}  \approx 5.4\times 10^{-20}$. 

 \begin{figure*}[!t]
		 \centering{\subfloat[$p=0.25$]{\includegraphics[width=0.5\columnwidth]{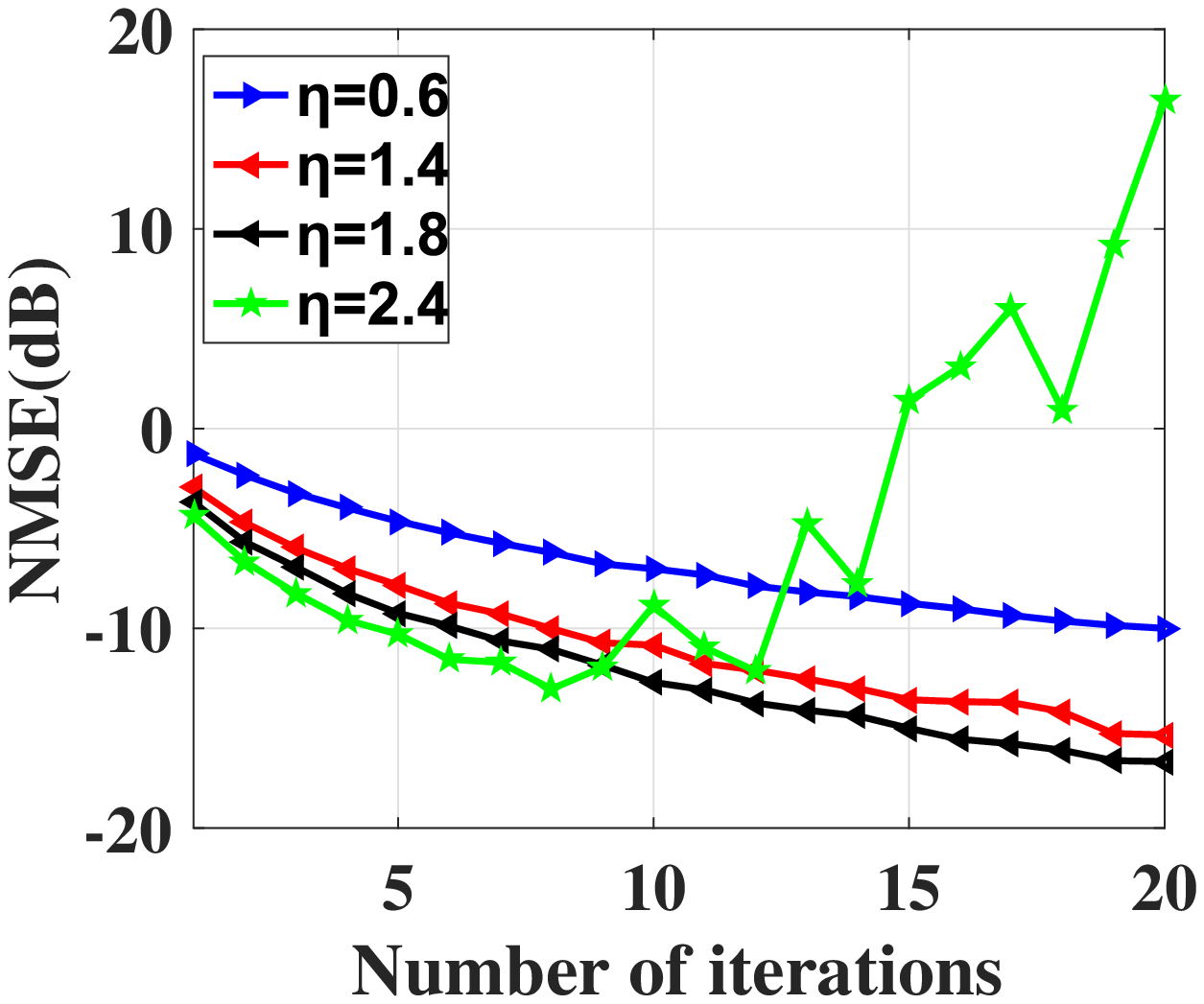}
		            \label{s1}}	           
	            \subfloat[$p=0.5$]{\includegraphics[width=0.5\columnwidth]{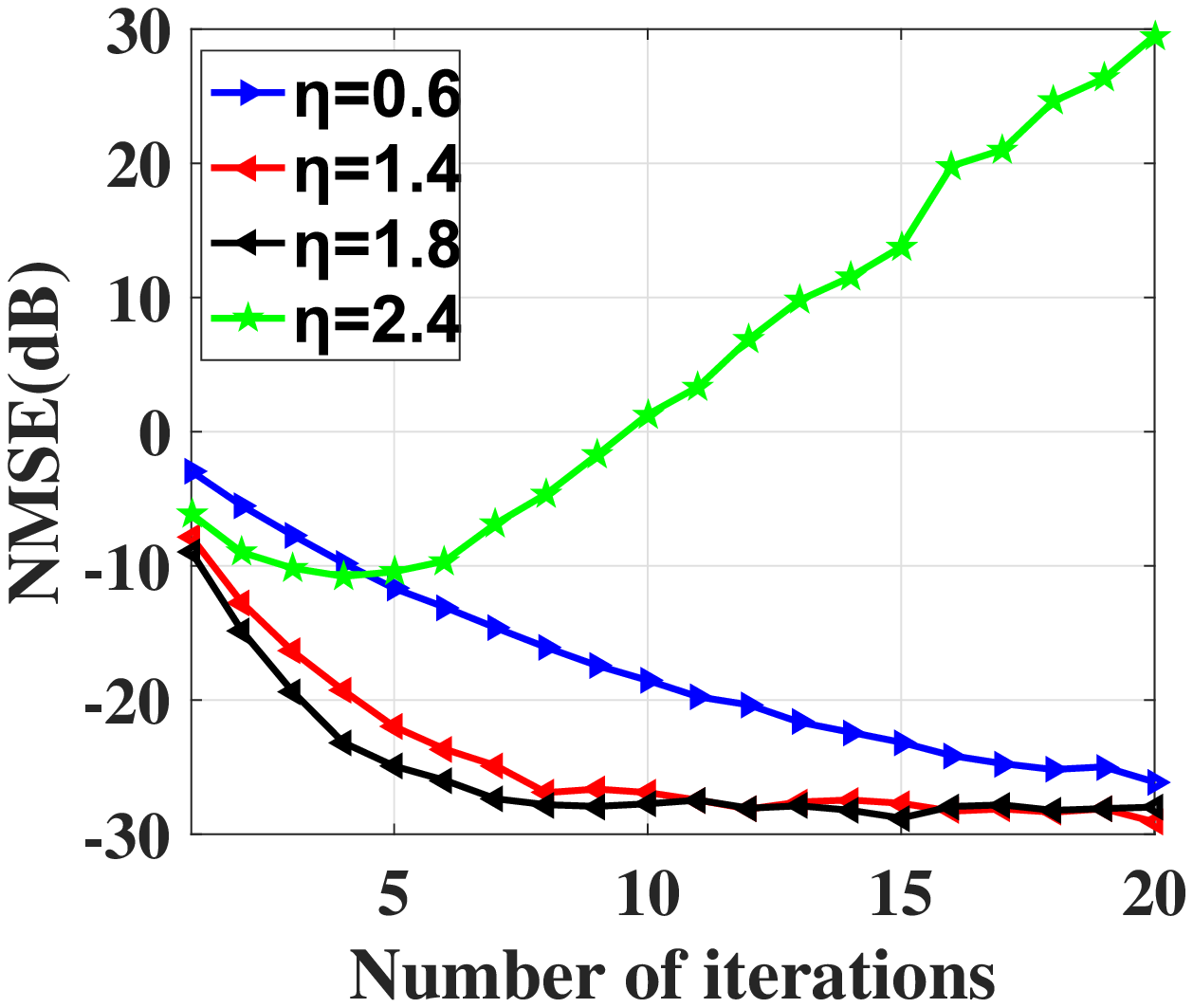}
	            \label{s2}}
\subfloat[$p=0.75$]{\includegraphics[width=0.48\columnwidth]{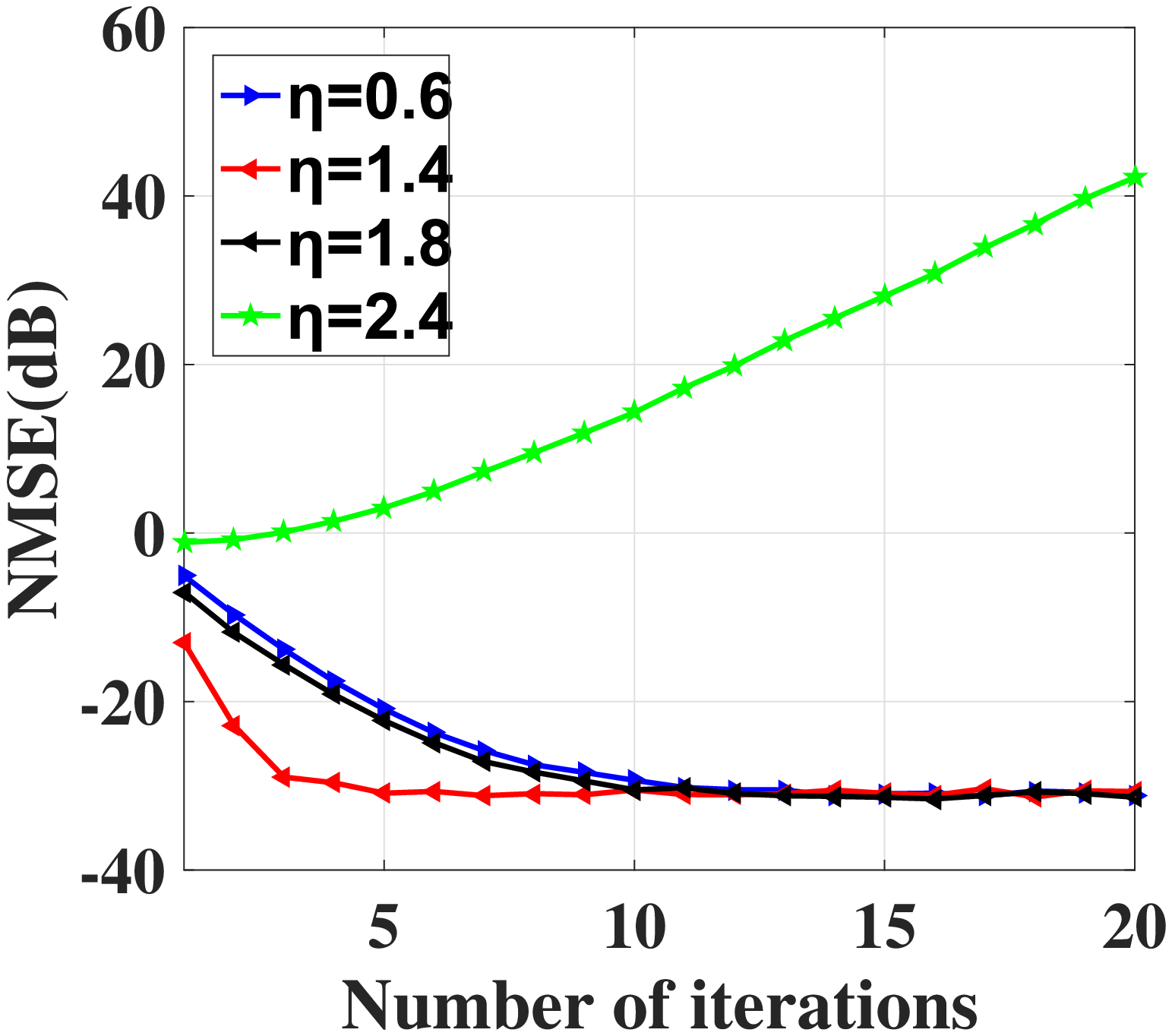}}
\subfloat[$p=0.5, \eta=1.8$]{\includegraphics[height=0.43\columnwidth, width=0.55\columnwidth]{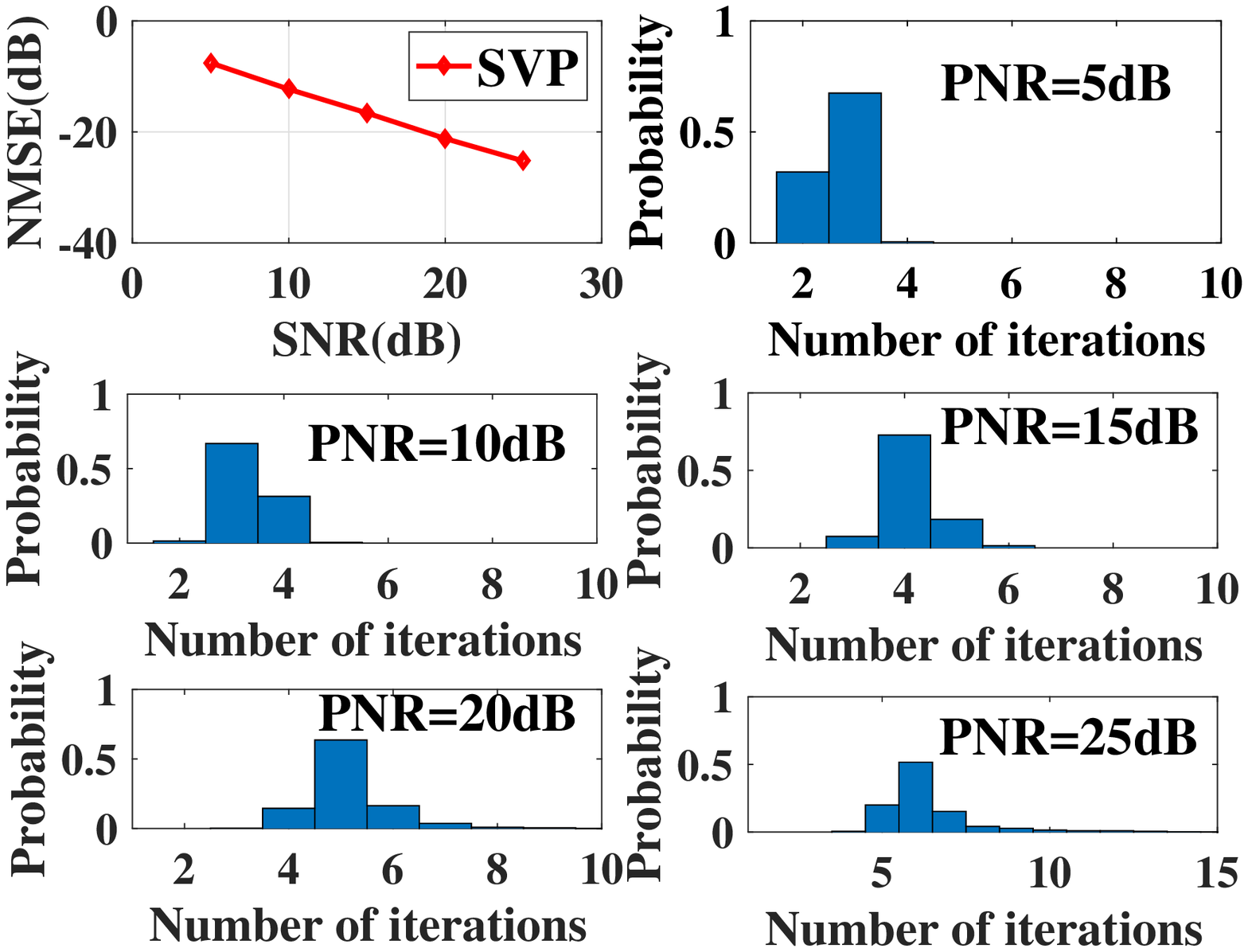}
	            \label{s2}}}
	    \caption{Choice of SVP parameters for the system where $N_{\rm BS}=N_{\rm {MS}}=64, L=4$. (a)-(c): Convergence performance with different levels of $\eta$ and $p$, PNR$=25$ dB; (d): Histograms of number of iterations to stop where $p=0.5, \eta=1.8, \rm{PNR}=5,10,15,20,25$ dB.	   }
	     \label{rate comparison}
\vspace{-4.5ex}
	 \end{figure*}

\begin{figure*}
		 \centering{\subfloat[Without phase mismatch]{\includegraphics[height=0.43\columnwidth, width=0.5\columnwidth]{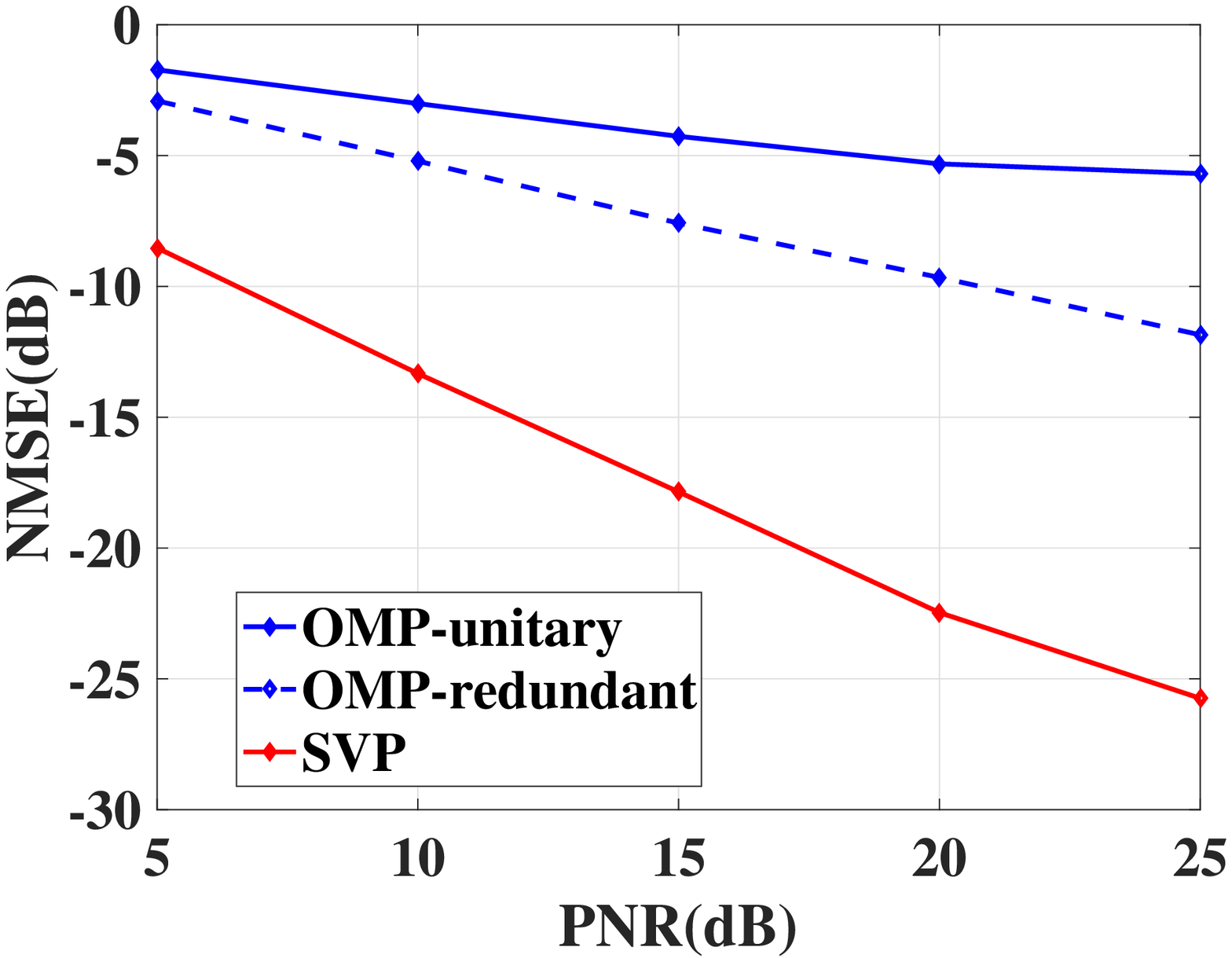}
		            \label{s1}}	           
	            \subfloat[With phase mismatch]{\includegraphics[height=0.43\columnwidth, width=0.51\columnwidth]{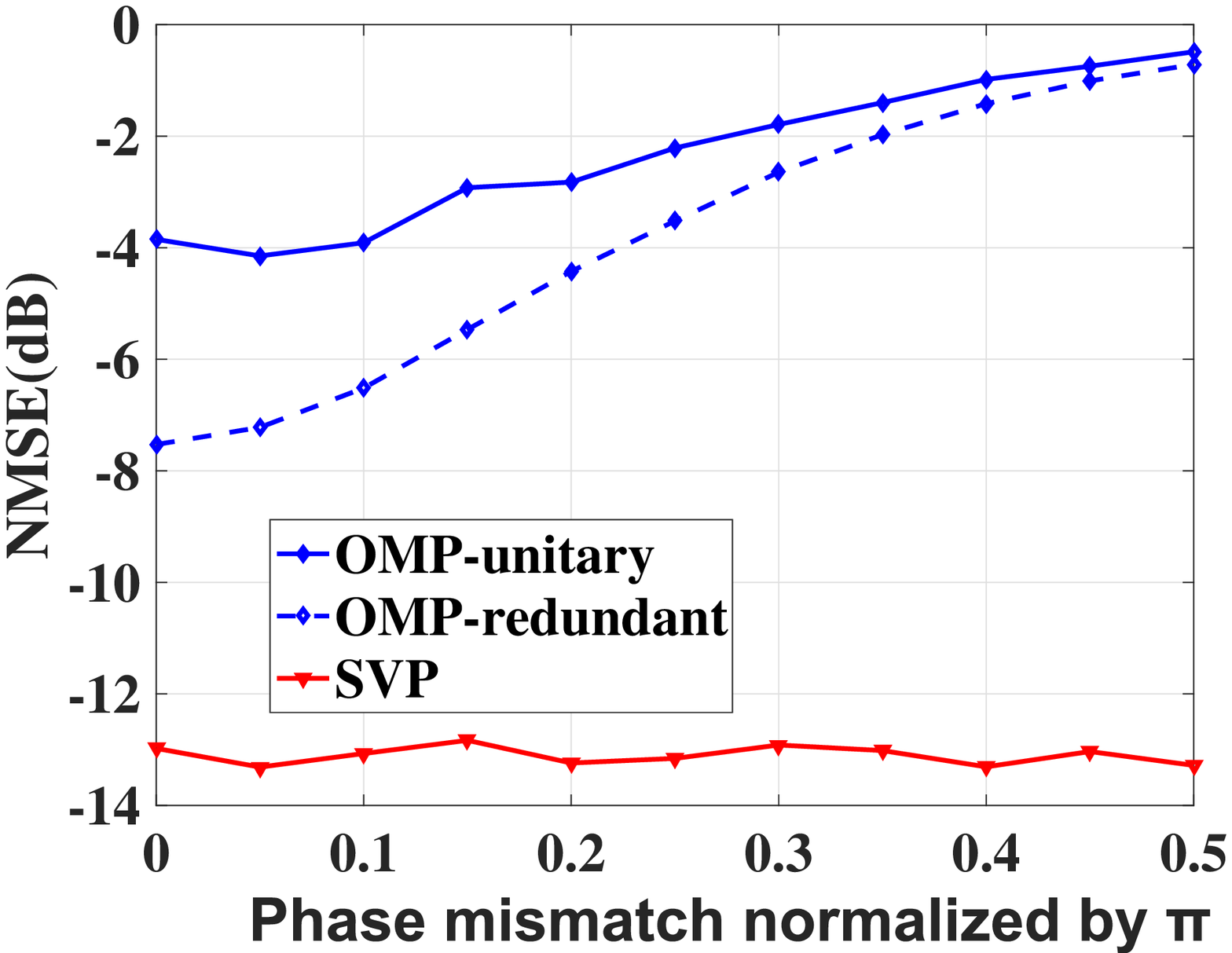}
	            \label{s2}}
                    \subfloat[ $\mathit{ Setting \; A}$]{\includegraphics[width=0.5\columnwidth]{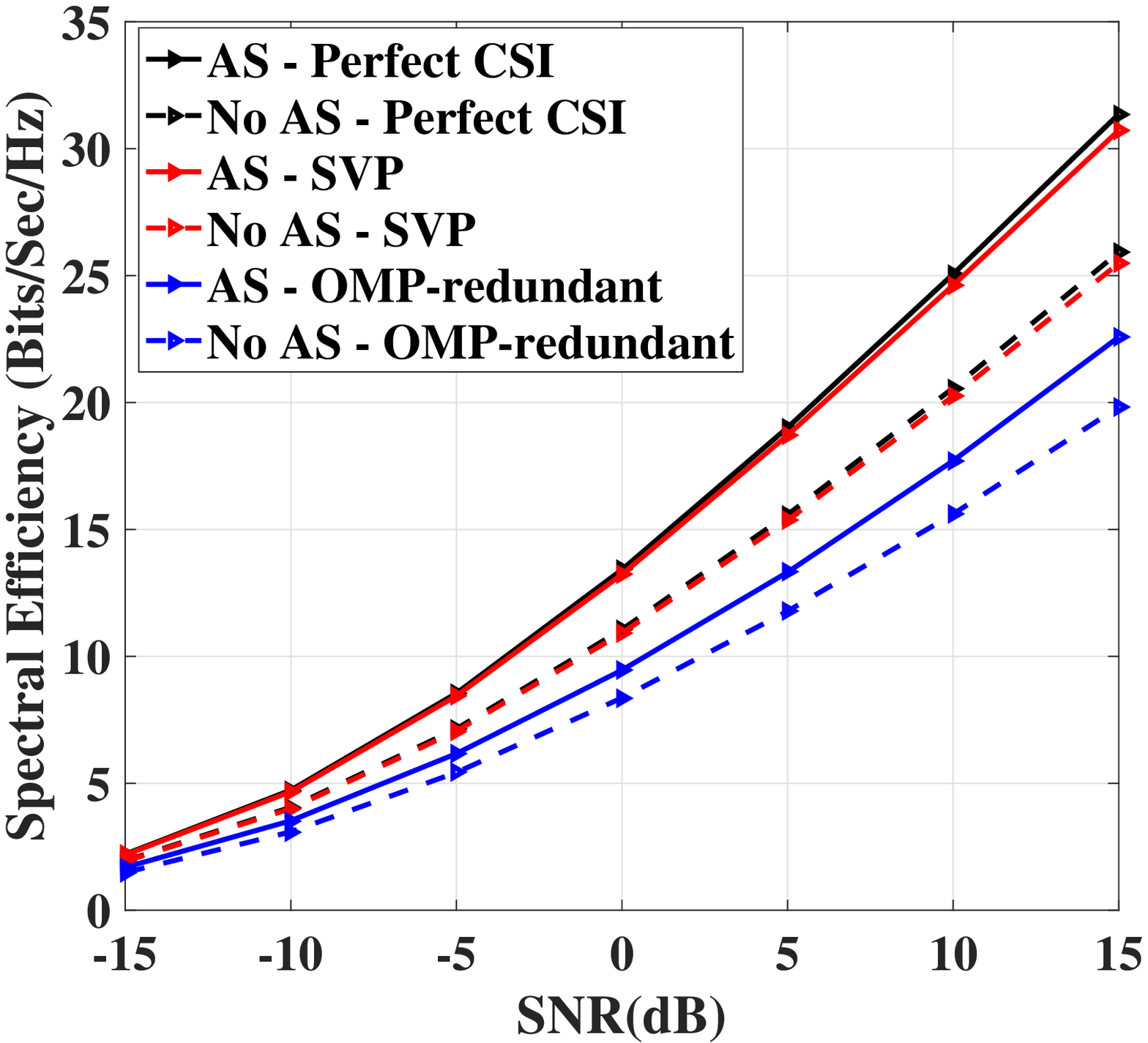}
	            \label{s2}}
 \subfloat[ $\mathit{ Setting \; B}$]{\includegraphics[width=0.5\columnwidth]{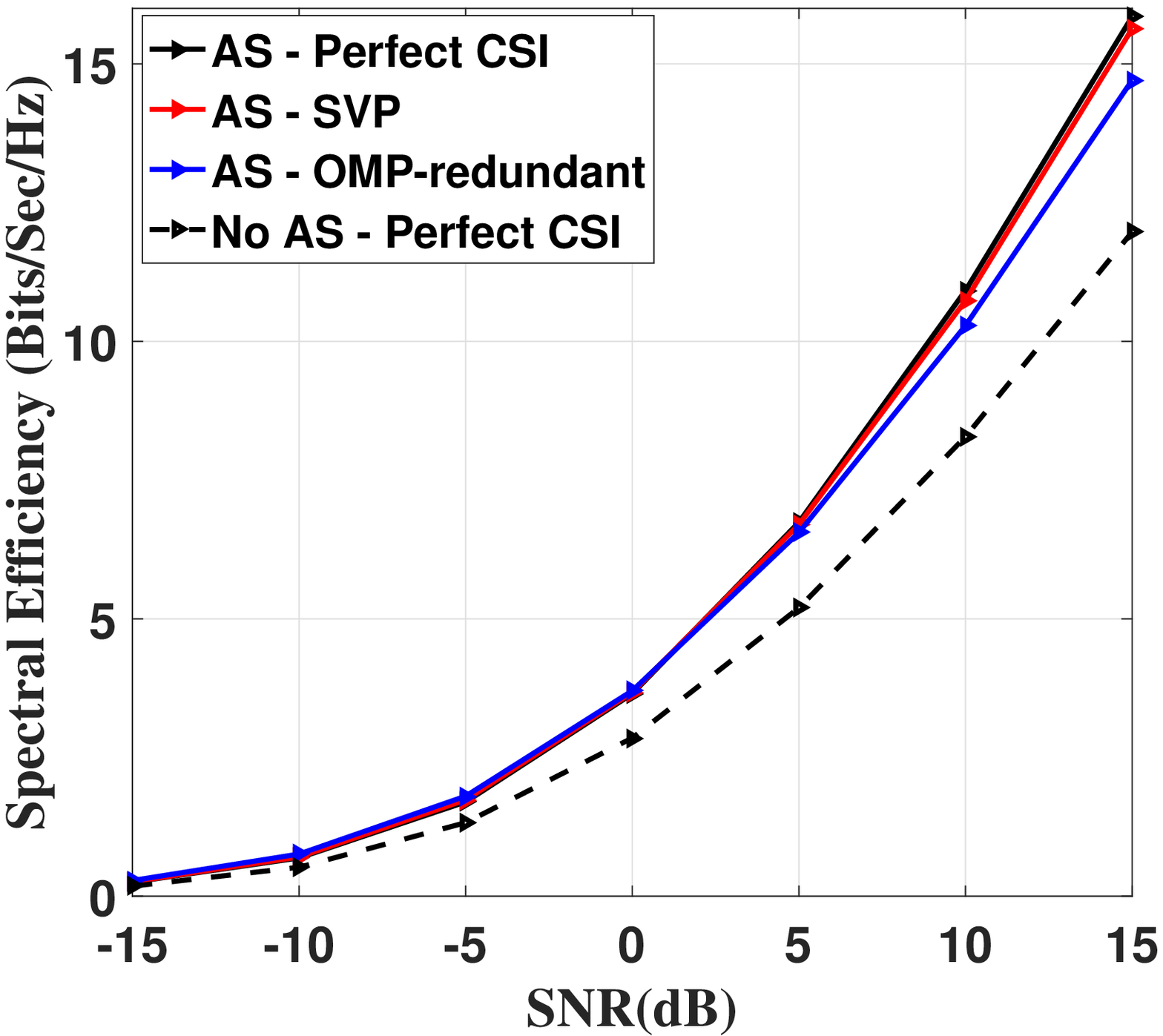}
	            \label{s2}}}
	    \caption{Comparison between the OMP estimator and the SVP estimator for the system where $ N_{\rm {BS}}=N_{\rm{MS}}=64, L=4, p=0.5, \eta=1.8$. (a): NMSE comparison without phase mismatch; (b): NMSE comparison with phase mismatch; (c): SE comparison for $\mathit{ Setting \; A}$ with $N_{\rm{RF_{\rm{MS}}}}=4$; (d): SE comparison for $\mathit{ Setting \; B}$ with $N_{\rm{RF_{\rm{BS}}}}=N_{\rm{RF_{\rm{MS}}}}=4$. }	 
	     \label{SCHistEigs}
\vspace{-2.5ex}
	 \end{figure*}
\vspace{-2ex}
\subsection{Singular Value Projection (SVP)}
After the training process, we apply singular value projection (SVP) algorithm \cite{SVP} to reconstruct $\mb H$. The algorithm is to solve the following matrix sensing problem
\begin{equation}
\label{matrix sensing}
\min_{\mb X} \psi (\mb X) = \frac{1}{2}\|\mathcal{A}(\mb X)-\bm b\|_F^2, \; \mathrm{s.t.}\; {\rm rank} (\mb X) \leq L, 
\end{equation} 
where $\mathcal{A}$ is a linear map, $\bm b$ is the observed signal and $L$ is the maximum rank of the matrix $\mb X$. The MC problem is a special case of the above matrix sensing problem, which replaces the sensing operator $\mathcal{A}$ by the operator $P_{\Omega}$ defined in (\ref{operator}). Therefore the problem becomes  
\begin{equation}
\label{matrix completion}
\min_{\mb X} \psi (\mb X) = \frac{1}{2}\|P_{\Omega}(\mb X)-P_{\Omega}({\mb Y})\|_F^2, \; \mathrm{s.t.}\; {\rm rank} (\mb X) \leq L, 
\end{equation}
where $\Omega$ characterizes the sampling pattern. 
A similar algorithm has been applied to MIMO channel estimation in a different scenario in \cite{Joint CSIT MC}. The SVP algorithm for solving the MC problem is shown in Algorithm I.
The major computational cost of the SVP is in Step 4, which needs to compute the rank-$L$ approximation of a $N_{\rm MS} \times  N_{\rm BS}$ intermediate matrix $\mb Z$. This can be done by computing the SVD of $\mb Z$. In order to reduce the high computational complexity due to SVD, we can choose an alternative way to calculate the rank-$L$ approximation of $\mb Z$ by first computing the eigenvalue decomposition of the $N_{\rm MS} \times  N_{\rm MS}$ matrix $\mb Z\mb Z^H=\mb U\mb S\mb U^H$, and then obtaining $\mb Z_L=\mb U_L\mb U_L^H\mb Z$, where $\mb U_L$ consists of columns in $\mb U$ that correspond to the $L$ largest eigenvalues. This way the computational cost of the  SVP algorithm is about $16N_{\rm MS}^2N_{\rm BS}+23N_{\rm MS}^3+8N_{\rm MS}^2L$ flops per iteration. 

\begin{table} 
\centering \small 
\begin{tabular}{ll} \hline
\textbf{Algorithm 1}  Singular Value Projection (SVP)\\ \hline 
\textbf{Input:}  $P_{\Omega}(\mb Y), L, \eta,\epsilon$ \\
\textbf{Initialization:}  $\mb X^0=\bm 0, t=0$ \\ 
1. \textbf{repeat} \\ 
2. $\mb Z^{t+1}\leftarrow \mb X^t-\eta(P_{\Omega}(\mb X^t)-P_{\Omega}(\mb Y)) $\\ 
3. Compute the top $L$ singular vectors of $\mb Z^{t+1}$: $\mb U_L,\mb \Sigma_L, \mb V_L$\\
4. $\mb X^{t+1}\leftarrow \mb U_L\mb \Sigma_L\mb V_L^H$\\ 
5. $t=t+1$\\
6. Until $\| P_{\Omega}(\mb X^t)-P_{\Omega}(\mb Y)\|_2^2\leq\epsilon$\\
\textbf{Output} the channel estimate  $\widehat{\mb H} = \mb X^{t}$\\ \hline  \\ 
\end{tabular}
\label{SVP} 
\end{table}

The convergence of the iterative SVP algorithm is 
influenced by the step size $\eta$. 
A small step size guarantees convergence but has low convergence rate, while a large step size implies fast convergence yet has the risk of divergence.
The authors of \cite{SVP} analyzed the convergence condition of the SVP algorithm for solving the general matrix sensing problem in (\ref{matrix sensing}) and suggested to set the step size $\eta=1/(1+\delta)<1$ with $\delta<1/3$, where $\delta$ is the RIP constant of linear map $\mathcal{A}$. For the MC problem, a special case of matrix sensing, the authors revised the step size to $\eta=1/(p(1+\delta))$ with $0<p<1$ and $\delta<1/3$.  This indicates that the step size can be larger than 1, e.g., for $p=0.25, \delta=1/3,\eta=3$. However, based on our observations, the step size can not be too large, e.g. if $p=0.25$, set $\eta=2.4$, the SVP method may diverge. Meanwhile, setting $1<\eta<2$ can obtain fast convergence rate. 
\vspace{-2ex}
\section{Simulation Results}
\subsection{Choice of SVP Parameters}
We first show the influence of $\eta$ on the convergence rate of the SVP estimator. 
 We assume that there are $N_{\rm{BS}}=64$ BS antennas and $N_{\rm{MS}}=64$ MS antennas, and there are $N_{\rm RF_{\rm MS}}=N_{\rm RF_{\rm BS}} = 4$ RF chains at the MS and the BS, respectively. The number of paths, i.e., the rank of the channel matrix, is $L=4$. The pilot-to-noise ratio (PNR) is 25dB. The AoAs and AoDs of $\mb H$ are uniformly distributed in $[ {-\pi}/{2}, {\pi}/{2}]$, and $\sigma_\alpha^2$ is set to 1. 
We use the normalized mean squared error (NMSE) defined as $ {\|\mb H-\widehat{\mb H}\|^2_F}/{\|\mb H\|^2_F}$ to evaluate the performance of SVP. Under this system setting and using our proposed sampling scheme, Fig.2 (a)-(c) show the convergence behaviour for different $\eta$ with  sampling density $p=0.25, 0.5, 0.75$, respectively.
 {It can be seen from Fig. 2 (a)-(c) that the SVP converges with $\eta=0.6,1.4,1.8$ for all the three cases, and faster convergence occurred when $\eta>1$. When $\eta=2.4$, the SVP diverges. 
From our simulation studies we also observed that the trends are similar for other levels of PNR and the convergence is faster when the PNR is lower. 

Based on the convergence analysis in \cite{SVP}, the tolerance $\epsilon$ of Algorithm I can be set as $\epsilon= C\|\mb e\|^2_F+\epsilon_0$, where $\|\mb e\|^2_F$ is the instantaneous total noise power of the observed entries, $C$ and $\epsilon_0$ are constants. Since $\|\mb e\|^2_F$ is unknown, we use $pN_{\rm{BS}}N_{\rm{MS}}\sigma^2$ to approximate $C\|\mb e\|^2_F$, where $\sigma^2$ is the average noise power. Fig. 2 (d) shows the stopping performance of using this tolerance $\epsilon$ for a system with $N_{\rm{BS}}=N_{\rm{MS}}=64, L=4, p=0.5, \eta=1.8$, and $\epsilon_0=10^{-3}$. As shown in the histogram in Fig. 2 (d), the convergence rate is different for different PNRs. For $\mathrm{PNR}=5,10,15,20,25$ dB, it takes $3,3,4,5,6$ iterations on average for the SVP to stop respectively. The histograms also indicate that the SVP method can stop within a small number of iterations by using the tolerance $\epsilon$.
\vspace{-3.5ex}
\subsection{Comparison of NMSE and SE}
We next compare the performance and computational complexity between the proposed SVP estimator and the OMP estimator discussed in Section II.
Assume $N_{\rm BS}=N_{\rm MS}=64$, $N_{\rm RF_{\rm MS}}=N_{\rm RF_{\rm BS}} = 4, L=4, p=0.5$. 
Two dictionaries are considered for the OMP. One is unitary with $G_t=64, G_r=64$ and the other is redundant with $G_t=128, G_r=128$. For the SVP, we set the step size $\eta=1.8$. The per-iteration complexity ratio between the OMP estimator with unitary basis and the SVP estimator is around $8pN_{\rm{BS}}N_{\rm{MS}}G_tG_r/(16N_{\rm MS}^2N_{\rm BS}+23N_{\rm MS}^3+8N_{\rm MS}^2L) \approx 6.5$, and the ratio increases to 26 with the redundant OMP basis. 
Both the OMP and SVP algorithms are iterative. 
We set the number of iterations to be the same for the OMP and SVP estimators,  
which is 2 for $\rm{PNR}= 5$ dB, 3 for $\rm {PNR}= 10$ dB, 4 for $\rm {PNR}= 15$ dB, 5 for $\rm {PNR}= 20$ dB and 6 for $\rm {PNR}= 25$ dB. The total computational complexity of the SVP scheme is about $1/6.5$ and $1/26$, respectively, of that of the OMP-unitary and OMP-redundant. 
From Fig.3 (a), the SVP estimator can outperform the OMP estimator at a much lower computational complexity. 

In practice, there can be phase mismatch among array elements. 
The OMP estimator that depends on the basis is sensitive to such array uncertainty. By contrast, the basis-free SVP estimator 
is immune to the phase mismatch. This is shown in Fig. 3 (b), where the unknown phase error is assumed to be uniformly distributed as $\gamma\sim U[-\gamma_{\rm{max}}, \gamma_{\rm{max}}]$ and 11 different levels of phase mismatch are considered by setting  $\gamma_{\rm{max}}=\{0,0.05\pi,\cdots, 
0.5\pi\}$. 

We further show the impact of the channel estimation scheme on the achievable spectral efficiency (SE) for MIMO transmissions employing the switch-based array structure under two different settings:  



$\mathit{ Setting \; A}$: Following \cite{switches or phase shifters}, the BS employs a fully digital beamformer with one RF chain equipped for each antenna, and the MS adopts the switch-based hybrid structure in Fig.1. 
During channel estimation, only one transmit antenna is activated to send the pilot at each training stage. Using the estimated channel $\widehat{\mb H}$, the BS precoder is chosen as $\mb P=\mb V_{\rm{L}}$, where $\mb V_{\rm{L}}$ consists of the $L$ dominant right singular vectors of $\widehat{\mb{H}}$. The incremental successive selection algorithm (ISSA) \cite{ISSA} is adapted to select $N_{\rm{RF_{\rm{MS}}}}$ out of $N_{\rm{MS}}$ MS antennas which can maximize the SE, with one antenna from each sub-array.

$\mathit{ Setting\; B}$: Both the BS and MS adopt the switch-based hybrid structure in Fig.1. We use the joint transmit-receive selection method in \cite{ISSA} to select $N_{\rm{RF_{\rm{BS}}}}$ out of $N_{\rm{BS}}$ BS antennas and $N_{\rm{RF_{\rm{MS}}}}$ out of $N_{\rm{MS}}$ MS antennas.

Fig.3 (c) and (d) show the impact of the antenna selection (AS) and channel estimation scheme 
on the achievable SE for $\mathit{ Setting \; A}$ and $\mathit{ Setting \; B}$, respectively. 
The channel estimation schemes are the same as those for Fig.3 (a), with a fixed $\rm{PNR=10}$ dB, similarly to the setting in \cite{channel estimation via OMP}.  
The results with ``No AS" are obtained by assuming a fully digital array with the number of antennas exactly equal to $N_{\rm{RF_{\rm{MS}}}}=4$ or $N_{\rm{RF_{\rm{BS}}}}=4$.    
From Fig.3 (c) and (d), employing larger antenna arrays with AS leads to significant gains compared to using smaller fully digital arrays. 
Furthermore, the SVP channel estimator leads to SE very close to that with perfect CSI while the OMP estimator, which exhibits a much higher complexity, results in noticeable losses in SE, especially for \emph{Setting A} in Fig.3 (c). 
This indicates that the proposed SVP estimator provides sufficiently good channel estimation at a much lower complexity than the OMP approach.  

\vspace{-1.5em}
\section{Conclusions}
In this paper, we show that matrix completion can be used for mmWave channel estimation. An estimation scheme that is compatible with the switch-based hardware structure is proposed. We show that the SVP method can exhibit significantly lower complexity than the OMP scheme and is immune to the phase mismatch of the array. Furthermore, we evaluate the impact of the channel estimation error on the achievable spectral efficiency (SE) for two different systems. The numerical results suggest that the SVP estimator can achieve near-optimal performance.

\vspace{-0.5em}


\end{document}